%% file: camera-ready.tex
\documentclass[sigconf]{acmart}
\AtBeginDocument{%
  \providecommand\BibTeX{{%
    \normalfont B\kern-0.5em{\scshape i\kern-0.25em b}\kern-0.8em\TeX}}}

\copyrightyear{2026}
\acmYear{2026}
\setcopyright{cc}
\setcctype{by-nc-nd}
\acmConference[CHI '26]{Proceedings of the 2026 CHI Conference on Human Factors in Computing Systems}{April 13--17, 2026}{Barcelona, Spain}
\acmBooktitle{Proceedings of the 2026 CHI Conference on Human Factors in Computing Systems (CHI '26), April 13--17, 2026, Barcelona, Spain}
\acmPrice{}
\acmDOI{10.1145/3772318.3790609}
\acmISBN{979-8-4007-2278-3/2026/04}
\usepackage{multirow}

\begin{document}

\title{Capability at a Glance: Design Guidelines for Intuitive Avatars Communicating Augmented Actions in Virtual Reality}


\author{Yang Lu}
\authornote{This work was done while Yang Lu was a visiting student at University of Rochester.}
\affiliation{%
  \department{College of Computer Science and Technology}
  \institution{Zhejiang University}
  \city{Hangzhou}
  \state{Zhejiang}
  \country{China}}
\affiliation{%
  \department{Department of Computer Science}
  \institution{University of Rochester}
  \city{Rochester}
  \state{New York}
  \country{USA}}  
\email{danaluyang@zju.edu.cn}

\author{Tianyu Zhang}
\affiliation{%
  \department{Department of Computer Science}
  \institution{University of Rochester}
  \city{Rochester}
  \state{New York}
  \country{USA}}  
\email{tianyu.zhang@rochester.edu}

\author{Jiamu Tang}
\affiliation{%
  \department{Goergen Institute for Data Science and Artificial Intelligence}
  \institution{University of Rochester}
  \city{Rochester}
  \state{New York}
  \country{USA}}  
\email{jtang41@u.rochester.edu}

\author{Yanna Lin}
\affiliation{%
  \department{Department of Computer Science and Engineering}
  \institution{The Hong Kong University of Science and Technology}
  \city{Hong Kong}
  \country{China}}  
\email{ylindg@connect.ust.hk}

\author{Jiankun Yang}
\affiliation{%
  \department{Department of Computer Science}
  \institution{University of Rochester}
  \city{Rochester}
  \state{New York}
  \country{USA}}  
\email{jyang118@u.rochester.edu}

\author{Longyu Zhang}
\affiliation{%
  \department{College of Computer Science and Technology}
  \institution{Zhejiang University}
  \city{Hangzhou}
  \state{Zhejiang}
  \country{China}}
\email{zhanglongyu@zju.edu.cn}

\author{Shijian Luo}
\authornote{Corresponding author}
\affiliation{%
  \department{College of Computer Science and Technology}
  \institution{Zhejiang University}
  \city{Hangzhou}
  \state{Zhejiang}
  \country{China}}
\email{sjluo@zju.edu.cn}

\author{Yukang Yan}
\affiliation{%
  \department{Department of Computer Science}
  \institution{University of Rochester}
  \city{Rochester}
  \state{New York}
  \country{USA}}  
\email{yukang.yan@rochester.edu}

\renewcommand{\shortauthors}{Lu et al.}

\begin{abstract}
Virtual Reality (VR) enables users to engage with capabilities beyond human limitations, but it is not always obvious how to trigger these capabilities. Taking the lens of \textit{Affordance}~\cite{norman1999affordance}, we believe avatar design is the key to solving this issue, which ideally should communicate its capabilities and how to activate them. To understand the current practice, we selected eight capabilities across four categories and invited twelve professional designers to design avatars that communicate the capabilities and their corresponding interactions. From the resulting designs, we formed 16 guidelines to provide general and category-specific recommendations. Then, we validated these guidelines by letting two groups of twelve participants design avatars with and without guidelines. Participants rated the guidelines’ clarity and usefulness highly. External judges confirmed that avatars designed with the guidelines were more intuitive in conveying the capabilities and interaction methods. Finally, we demonstrated the applicability of the guidelines in avatar design for four VR applications.
\end{abstract}




\begin{teaserfigure}
  \includegraphics[width=\textwidth]{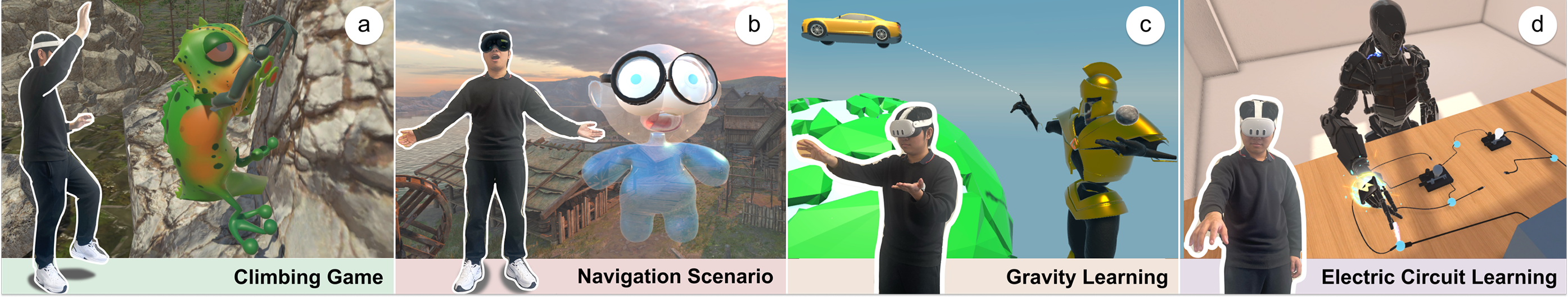}
  \caption{Avatars designed with our design guidelines implemented in four VR applications: 
  (a) the enhanced climbing avatar encourages full-body engagement in a climbing game; 
  (b) the body size manipulation avatar supports intuitive and controllable navigation speed adjustment; 
  (c) the gravity manipulation avatar facilitates a conceptual understanding of gravity through embodied learning; 
  (d) the electricity manipulation avatar enables hands-on and safer interaction in circuit learning tasks.
  }
  \label{fig:teaser}
  \Description{Teaser}
\end{teaserfigure}


\maketitle
\input{sections/01_introduction}

\input{sections/02_related_work}
\input{sections/03_formative_study}

\input{sections/04_Guidelines}
\input{sections/05_Evaluation}
\input{sections/06_Implementation}

\input{sections/07_Discussion}
\input{sections/08_Conclusion}
\begin{acks}
The authors thank the reviewers for their constructive suggestions, all participants for their time and effort, and the Bear Lab members for their valuable support and discussions.
\end{acks}

\bibliographystyle{ACM-Reference-Format}
\bibliography{sample-base}

% \appendix









\end{document}

%% file: sections/01_introduction.tex
\section{Introduction}
In Virtual Reality (VR), users can embody avatars that are totally different from their physical bodies~\cite{jiang2023handavatar} and become free from the constraints of the real world (e.g., by breaking the laws of physics)~\cite{sadeghian2021limitations}.
Augmented actions that exceed human capabilities, such as flying~\cite{zhang2019perch} and extending reach~\cite{lin2024armdeformation}, can be enabled to complete different tasks, therefore benefiting various application scenarios, including education~\cite{pirker2021potential,jensen2018review}, well-being~\cite{riva2002virtual}, and entertainment~\cite{dey2017effects}. 
Significant progress has been made in developing interaction techniques for performing augmented actions in VR.
For example, foot-based gestures (e.g., point and lean)~\cite{von2020podoportation} were designed to support teleportation (an augmented action that moves the avatar over long distances immediately).
Although powerful, users can overlook these augmented actions as they do not expect to have the capabilities or know how to trigger them. 
We argue that it is essential to make these capabilities self-revealing to the users, to improve the adoption of the interaction techniques, and further release the potential of VR in target applications.

We argue that avatar design offers a promising but underexplored opportunity to address this gap. 
Avatars play a central role in shaping user perception and behavior and forming affordances to indicate beyond-human capabilities~\cite{oberdorfer2024proteus,ogawa2020you,pena2017great,dufresne2024understanding,bhargava2023can}. 
We adopt Norman’s definition of affordances~\cite{norman1995psychopathology,norman1999affordance} as the perceived action possibilities available to the user, allowing them to understand what to do without explicit instructions. 
In this context, leveraging affordances through the visual aspects of avatars can effectively convey the augmented capabilities and the corresponding interaction methods~\cite{seinfeld2021user}.  
For example, when users see their avatar represented with wings instead of arms, they may naturally associate this visual cue with the capability to fly. 
Applying an additional design as the arm-wing replacement, they may easily realize the interaction method to trigger it, which is to flap their arms to trigger the flying action. 
However,  in current applications, avatar design is not always effective in conveying the information, as the design process typically relies on individual designers’ experience, intuition, and creativity without a shared guideline. 

In this sense, \textbf{we propose to develop a set of avatar design guidelines to systematically guide designers in creating virtual avatars that can effectively communicate augmented actions to users.} 
To achieve the goal, we began by understanding how expert designers create avatars to communicate augmented actions. 
We classified common augmented capabilities into four main categories based on prior studies\cite{chodos2014framework,fardinpour2014taxonomy,willett2021perception} and our synthesis of capability characteristics. 
We then selected 12 representative instances, three from each category, and invited 12 professional designers to create avatars for them, resulting in 27 avatar designs. 
Through analyzing the results, we summarized key design strategies and formed a set of 16 avatar design guidelines. 
To evaluate our guidelines, we conducted a user study where two groups of twelve participants designed avatars with and without guidelines respectively, generating 25 avatar designs~\footnote{P8 created two avatar designs for freezing.}. 
We invited 48 external judges to assess the intuitiveness of these resulting avatars through an online survey. 
Participants who designed avatars with guidelines rated clarity and usefulness of the guidelines highly in ideation and design processes, with average scores above 5 on a seven-point Likert scale. 
The evaluation results showed that the avatars designed with guidelines were more intuitive in communicating augmented capabilities.
Finally, to demonstrate the applicability of the guidelines, we implemented four avatars designed with guidelines in four VR applications in sports games, navigation, and education scenarios.
We conducted a demonstration study where 12 participants experienced the four applications using associated avatars. 
They reported that the intuitive avatar designs were easy to understand and enhanced their sense of immersion, engagement, and enjoyment. 
In summary, our contributions are threefold: 
\begin{itemize}
    %\item We proposed a classification framework of augmented actions grounded in prior literature and the functional characteristics of such actions.
    \item We conducted a user study to understand the current practice of designers in designing avatars to communicate augmented capabilities. Based on the study results, we developed 16 avatar design guidelines that provide actionable instructions for different steps in the design process, from early-stage inspiration to detailed elements such as color, material, and visual effects. 
    \item We validated the effectiveness and practicality of the proposed guidelines through a second study, where we verified that avatars designed with the guidelines were more intuitive in conveying augmented actions than those without them.
    \item We implemented four avatars designed with the guidelines in VR and demonstrated their effectiveness in communicating the corresponding augmented actions in the education, well-being, and entertainment applications.

\end{itemize}

%% file: sections/02_related_work.tex
\section{Related Work}
\begin{figure*}
    \centering
    \includegraphics[width= 1\textwidth]{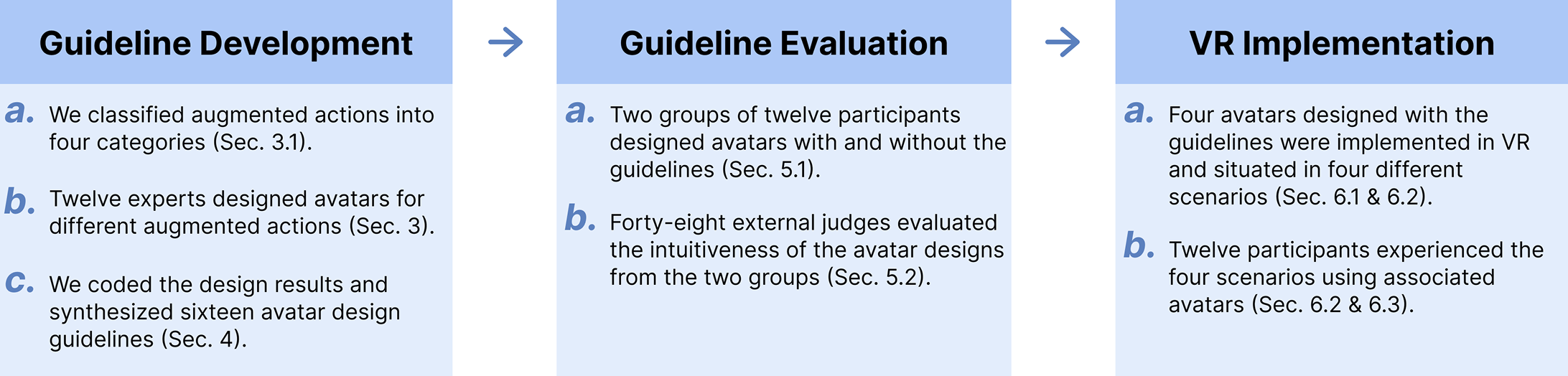}
    \caption{Overview of the paper structure. We followed a three-stage process: (1) Guideline Development: we classified augmented actions and conducted a design study with twelve experts to sythesize avatar design guideliens; (2) Guideline Evaluation: two groups of twelve participants designed avatars with and without guidelines, and forty-eight external judges evaluated their intuitiveness; and (3) VR Implementation: four avatars designed with guidelines were implemented in four VR scenarios and twelve participants experienced them in a user study.}
    \label{fig:frame}
\end{figure*}
\subsection{Augmented Action in VR}
Augmented action is a key aspect of human augmentation~\cite{guerrero2022augmented,alicea1804integrative,raisamo2019human}, which improves a human's ability to perform physical actions, including motor enhancement, increased strength, and remote manipulation~\cite{guerrero2022augmented,raisamo2019human,de2021structuring,de2024human}. 
Recent research highlights the potential of VR to go beyond mere replication of reality~\cite{mostafa2014poster,neuhaus2024mimic,sadeghian2021limitations}. 
Consequently, leveraging VR technology to implement augmented actions has become a significant research focus. 
Previous research has introduced various augmented actions in VR, including enhanced locomotion techniques (e.g., super leaps~\cite{sasaki2019virtual}, teleportation~\cite{bozgeyikli2016point}, flying~\cite{zhang2019perch}), remote object interactions (e.g., extended reach~\cite{lin2024armdeformation,mendes2017precious}, multi-selection~\cite{schjerlund2021ninja}), and body-related modifications (e.g., phasing through objects~\cite{dufresne2024understanding}). 

These actions often lack real-world counterparts, making them unfamiliar and non-intuitive for users. 
As they become increasingly common in VR, a key challenge lies in effectively communicating them to ensure intuitive perception and use. 
However, the most effective strategies for helping users understand and engage with such augmented actions still remain an open research question~\cite{abtahi2022beyond}. 
To explore a better solution, we began by developing a classification of augmented actions in VR. 
Currently, there is no established taxonomy, so creating a structured categorization could lay a clearer groundwork for design. 
This framework would help analyze various augmented actions and support the development of intuitive design strategies tailored to each category.

\subsection{Communicating Augmented Actions in VR} 
Current research on augmented actions in VR has explored different aspects of how these actions are communicated to users. 
One line of work focused on how users initiate and control augmented actions through natural and engaging interaction techniques~\cite{zhang2019perch,bozgeyikli2016point,schjerlund2021ninja}. 
For example, Zhang et al.~\cite{zhang2019perch} introduced a flexible perching stance to support flying locomotion control, increasing the feeling of flying. 
Another line of research explored how to convey the effects of augmented actions through feedback mechanisms, including visual, haptic, and multi-sensory cues~\cite{lin2024armdeformation,sasaki2019virtual}. 
For instance, Lin et al.~\cite{lin2024armdeformation} employed skin-stretch sensations to help users perceive arm deformation during extended reach. 
These approaches all contribute to communicating augmented actions, either during or after interaction. 

However, little attention has been paid to how users recognize these actions before engagement. 
To better understand this perceptual process, we define an augmented action as the combination of an augmented capability (what users can do beyond natural limits) and its corresponding interaction method (how to trigger that capability).
In this view, users must first understand both what they can do and how to trigger it in order to successfully perform the augmented action. 
This front-loaded phase of perception is essential for actions that lack real-world equivalents. 
To address this gap, we propose using intuitive avatar designs to communicate augmented capabilities and interaction methods in advance. 

\begin{table*}[h]
    \centering
    \caption{The classification of augmented actions is divided into four categories based on their target and purpose, with examples provided for each category.}
    \resizebox{\textwidth}{!}{
    \begin{tabular}{|c|c|c|}
    \toprule
    Target & Purpose & Example Actions \\
    \midrule
    \multirow{2}*{Self-control}& Physical Self-movement & Enhanced jumping, Enhanced swimming, Enhanced climbing, Flying, Teleportation.
 \\
    \cline{2-3}
     & Bodily Transformation & Body size manipulation, Invisibility, Intangibility, Self-cloning, Morphing.
 \\
    \hline
    \multirow{2}*{External-manipulation}& Operative Action & Enhanced lift, Telekinesis, Magnetism, Gravity manipulation, Force field manipulation.
 \\
    \cline{2-3}
     & Constructive Action & Mountain manipulation, Freezing, Petrification, Gold transmutation, Fire manipulation
\\
\bottomrule
    \end{tabular}}

    \label{tab:actionclassification}
\end{table*} 

\subsection{Avatar Design and Affordance}
In Virtual Reality Environments (VREs), users interact with the virtual world through avatars, which serve as embodied representations that mediate users' virtual actions~\cite{seinfeld2021user,davis2009avatars} and influence users' perception and behavior~\cite{10269053,6479188,9417780}. 
Researchers have drawn upon affordance theory to understand how avatars influence interaction within VREs~\cite{arend2021object,lin2015affordance}. 
Affordance was originally coined by Gibson~\cite{gibson2014ecological}, and later adopted by Norman~\cite{norman1995psychopathology} for Human-Computer Interaction (HCI) to describe the perceived action possibilities available to a user. 
Norman further noted that well-designed affordances allow users to intuitively understand what to do without the need for explicit instructions~\cite{norman2013design}. 
Conversely, hidden or misleading affordances can lead to confusion and errors~\cite{gaver1991technology}. 
Prior studies on avatar-based affordance have shown that changes in avatar features, such as altering virtual body size\cite{jun2015big,buche2018revam}, hand representation~\cite{joy2022trick,dufresne2024understanding,venkatakrishnan2023virtual}, or avatar appearance~\cite{ogawa2020you,bodenheimer2015effect}, significantly influence how users perceive action possibilities and interact with virtual environments. 
However, most of these works focused on users’ perceptions of affordances during daily tasks, such as grasping objects~\cite{joy2022trick}, where users can rely on real-world analogies to interpret. 
In contrast, conveying augmented actions, a lack of a real-world equivalent, is significantly challenging. 
A few recent works have hinted at the potential of avatars to communicate augmented actions. 
For example, Christou et al.~\cite{7012029} found that users embodied in a tough alien-like avatar were more inclined to block projectiles with their hands, as these avatars conveyed stronger affordances for resisting cannon fire compared to bare anthropomorphic hands. 
Dufresne et al.~\cite{dufresne2024understanding} found that a ghost-like hand was perceived to afford the possibility of passing through objects. 
However, these studies have primarily observed these effects rather than systematically exploring how to intentionally use avatar design to communicate augmented capabilities and interaction methods.

To fill this gap, we developed a set of design guidelines to help create avatars that effectively convey augmented actions and the corresponding interaction methods. 
We derived these guidelines from a formative study involving 12 experts. 
To validate their effectiveness, we conducted an empirical study assessing whether the guidelines facilitate the creation of intuitive avatars. 
Furthermore, we implemented four guideline-compliant avatars in VR and invited participants to experience and evaluate them across various scenarios, demonstrating their practical benefits in VR applications. 

\autoref{fig:frame} provides an overview of the development, evaluation, and VR implementation of our guidelines. 
In Section 3, we describe the process of guideline development, including the classification framework for augmented actions and the expert design study. 
Section 4 details the sixteen avatar design guidelines we synthesized from the study.
Section 5 presents the evaluation of our guidelines, where two groups of participants designed avatars either with or without access to our guidelines, followed by an evaluation of avatar intuitiveness by forty-eight external judges.
In Section 6, we demonstrate the implementation of four avatars designed with guidelines in VR. 
We situated the four avatars in different scenarios and conducted a user study where twelve participants experienced these scenarios using the associated avatars.

%% file: sections/03_formative_study.tex
\begin{table*}[!ht]
    \centering
    \caption{Background information of the 12 experts who participated in the formative study }
    \resizebox{\textwidth}{!}{
    \begin{tabular}{cccc}
    \toprule
    Participant ID & Gender & Professional Background & Experience \\
    \midrule
    E1 & F & Game design & 5 years of experience in the game environment and character design \\
    E2 & F & Game design & 5 years of experience in 3d character modeling and animation \\
    E3 & F & Interaction Design & 2 years of experience in character design and game interaction experience design  \\
    E4 & M & Game Art & 7 years of experience in 3D character art \\
    E5 & F & Game design & 6 years of experience in 3d character modeling \\
    E6 & F & Interaction design & 2 years of experience in game character design \\
    E7 & F & Game Design &  5 years of experience in game character design \\
    E8 & M & User Experience Design &  6 years of experience in character design, had experience in VR-based virtual guide avatar design project \\
    E9 & M & Manga &  4 years of experience in character design for manga and illustration\\
    E10 & F & User Interface Design & 4 years of experience in character design, designed five metaverse avatar characters for an automotive company \\
    E11 & F & Game Design &  5 years of experience in game character design\\
    E12 & F & Manga &  4 years of experience in character design for manga and illustration\\
\bottomrule
    \end{tabular}}

    \label{Expertsbg}
\end{table*}
\section{Study 1: Avatar Design for Augmented Actions}
We conducted an elicitation study to investigate how experts design avatar appearances to hint at augmented actions.
Twelve augmented actions were selected and tested.
To ensure that they are representative, we classified the augmented actions into four main categories and selected three instances from each category.
Based on the designs elicited, we formulate design guidelines that designers can refer to when creating avatars with different augmented actions. 

\subsection{Classification of Augmented Actions}
To ensure the generalizability of our elicitation results, we first developed a structured classification of augmented actions and then selected representative actions from each category for the study. 
Our classification specifically targets augmented actions that require users to actively and intentionally trigger them, including motor enhancement, enhanced manipulation of the self-body and external environment. 
We exclude abilities that are purely cognitive, passive, or system-driven without embodied activation, which do not require user learning and can be presented automatically. 
We classified augmented actions into 4 categories based on the prior literature and our summary of the augmented actions' characteristics. 
First, we collected and reviewed literature on augmented actions implemented in immersive environments, extracting relevant examples of such actions. To expand our dataset, we conducted online searches using the Fandom Superpower Wiki~\cite{superpower_wiki}, a comprehensive pop-culture encyclopedia with the largest collection of superpowers that has frequently been adopted in related research~\cite{willett2021perception}. 
From these sources, we extracted augmented actions suitable for VR applications. This process resulted in a curated list of 36 distinct augmented actions.

Building on this dataset and drawing inspiration from prior studies about the action classification in VR~\cite{chodos2014framework,fardinpour2014taxonomy} and the classification of superpowers~\cite{willett2021perception}, we preliminarily classified augmented action into four major categories.
First, we divided the actions based on the action target into two primary groups: those focused on the user’s own body (self-control) and those targeting the external environment or object(external manipulation). Next, we subdivided each group based on the purpose of the action, creating a four-category structure~(\autoref{tab:actionclassification}). 

\textbf{\textit{Physical Self-movement}} refers to actions that improve users' physical performance, such as speed or distance, or enable supernatural movements that are impossible in real life, including enhanced jumping, running, flexibility, and flying.

\textbf{\textit{Bodily Transformation}} refers to actions that allow individuals to modify their body's attributes, such as appearance, shape, or physical state. 
Examples include body size manipulation, becoming intangible, or creating self-clones.

\textbf{\textit{Operative Action}} enhances natural interactions with the environment by increasing users' physical strength, reach, and control over objects in VR without altering the nature or state of the external environment.  
For example, lifting heavy objects like buildings or cars, and manipulating objects using magnetism.

\textbf{\textit{Constructive Action}} refers to manipulating the environment or objects to create significant effects. It can involve creating, destroying, or altering various objects' shape, state, or material. 
This category also includes powerful magical capabilities, such as reshaping terrain, controlling natural elements, and influencing the weather. Examples include mountain manipulation, fire manipulation, and gold transmutation.
\begin{figure*}
    \centering
    \includegraphics[width=1\textwidth]{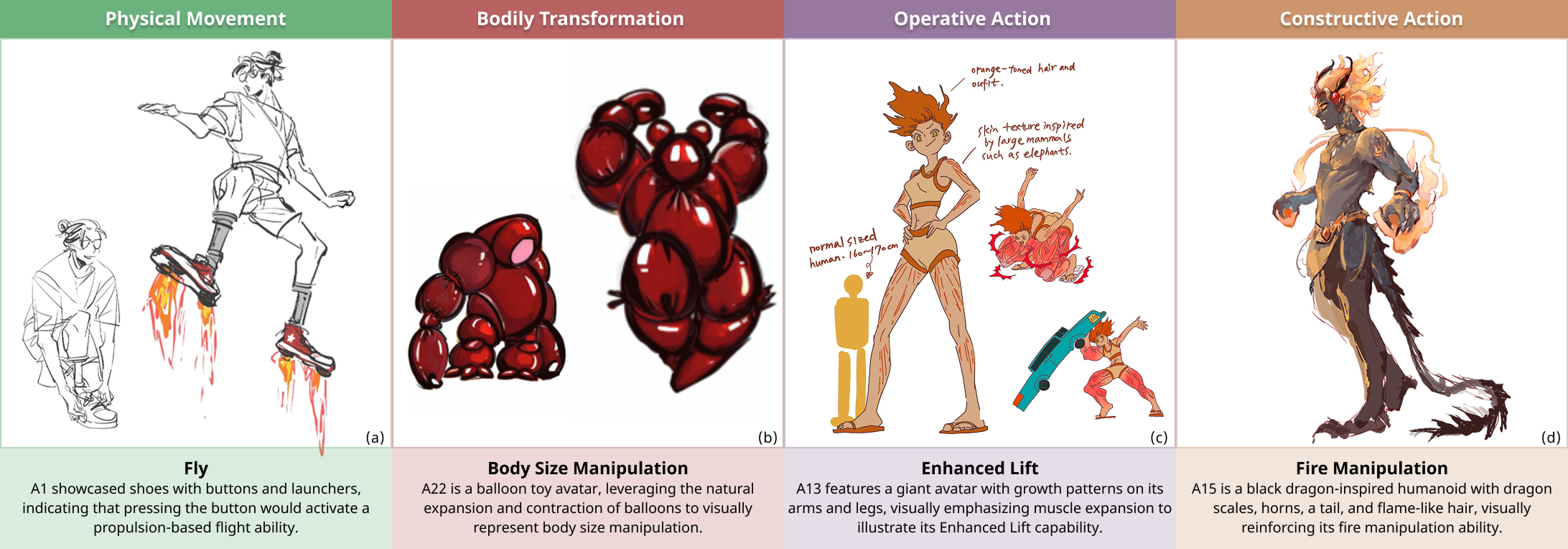}
    \caption{This figure showcases four avatars designed by experts in the formative study, each representing one of the four categories of augmented actions.}
    \label{fig:formativeoutcomes}
\end{figure*}
\subsection{Participants}
We recruited twelve expert designers (aged 23 to 29 years, M = 25.41, SD = 1.50; 3 male, 9 female) from the game industry and design institutions, all of whom had obtained a bachelor’s degree with design-related majors in interaction design, game design, user interface design, and manga (\autoref{Expertsbg}). 
They had completed formal coursework in areas such as character design, visual communication, and user-centered design. 
They had at least 2 years of experience in character design and intermediate proficiency-level illustration skills. 
Because they had experience visually conveying personality and abilities through character design, we believe that they were well-equipped to design avatars that clearly communicate augmented capabilities and how they are activated. 
Each participant received a compensation of \$60 for their participation. 
The study protocol was approved by the Institutional Review Board (IRB). 

\subsection{Task} 
To ensure comprehensive coverage of the augmented actions and the generalizability of our findings, we selected 12 representative actions, with 3 from each of the four categories. 
These actions were carefully selected to represent distinct aspects within each category. 
For example, in the constructive action category, we included mountain manipulation to represent terrain reshaping, fire manipulation for elemental control, and gold transmutation for material transformation. 
Details of the remaining augmented actions used in the study are provided in the supplementary materials. 
Due to time constraints, it was not feasible for each expert to create avatars for all 12 actions. 
However, we wanted to gather different interpretations of each action to compare and gain deeper insights. 
As a result, each expert was randomly assigned two augmented actions, ensuring that each action was designed twice by different experts. 
They were asked to design at least one corresponding avatar and interaction method for each action, with at least two avatar designs in total. 
Participants were instructed to design the visual aspects of avatars to clearly communicate their augmented capabilities and to indicate how these capabilities are activated. 
In doing this, users should be able to intuitively understand both the augmented capabilities and their activation methods by simply observing the avatar, without relying on textual explanations or external instructions. 
Participants were asked to manually design the avatars and were prohibited from using AI tools to generate designs, ensuring the design's originality. 
Participants were encouraged to explore multiple creative ideas. 
Finally, each expert submitted a design documentation package, which included avatar design sketches, design descriptions, and explanations of the triggering interaction. 

\subsection{Procedure} 
The study was conducted online, allowing participants to complete tasks independently and without supervision. 
This approach offered increased flexibility in scheduling and reduced potential stress. 
Before starting, each participant attended a one-on-one online session (about 10 minutes) where the experimenter explained the study procedures, augmented actions, and requirements. 
They also received a task description document and two example designs for reference. 
All materials were explained and discussed during the meeting to ensure consistent understanding, and participants were instructed to review them carefully before beginning their design. 
We maintained ongoing communication with the experts throughout the design process. Whenever they had questions, they could reach out to us, and we provided timely clarifications to ensure they understood the design objectives and task requirements correctly. Once prepared, participants provided their demographic information and completed the design tasks for two augmented actions in two two-hour sessions, which were spread over two days to avoid overloading. 
They were advised to work in a quiet, distraction-free environment to ensure focus and minimize interruptions. 
After completing the task, participants submitted their results. 
Finally, we collected 27 avatar design outcomes (A1-A27). 
\autoref{fig:formativeoutcomes} showcases four avatars designed by participants for four augmented actions from each category. 

\subsection{Analysis}
We aimed to identify common patterns in the avatars designed by experts for different categories of augmented actions. 
To achieve this, we coded their designs into multiple dimensions, including inspiration, avatar type, structure and form, outfits, visual effects, symbolism, material and texture, color, and interaction. 
Then, we compared them to elicit common patterns across all avatars and within specific categories.
Based on these results, we synthesized a set of design guidelines broadly applicable to avatar design for augmented actions.
In the coding process, the first and second authors analyzed the design outcomes. 
First, we independently extract and categorize design elements and features through an open coding approach~\cite{charmaz2006constructing}. 
Then, we had discussions to reach a consensus.
The design outcomes included both images and textual descriptions. We analyzed the design elements from the sketches, particularly when key visual features were not clearly described in the text. 
To ensure interpretative validity, we sent the coded files back to the experts for verification and feedback, ensuring our interpretations aligned with their original design intentions.
After that, the first, second, and third authors conducted brainstorming sessions using card sorting and affinity diagramming on an online whiteboard (Miro board). 
Through this process, we grouped design elements, identified underlying patterns, and refined these patterns into 16 structured design guidelines.

%% file: sections/04_Guidelines.tex
\section{Design Guidelines}
\begin{table*}[h]
\centering
\caption{Sixteen guidelines for intuitive avatars: eight general guidelines for all augmented actions and two specific guidelines for each of the four action categories.}
\resizebox{\textwidth}{!}{
\begin{tabular}{|c|c|c|}
\hline
       & \textcolor{blue}{General} & \\
   \hline
   1.1 & Draw inspiration from diverse sources, including nature, myths, everyday objects, and media, to create unique and intuitive avatar designs and interactions. & Recommended \\
   \hline
   1.2 & Consider semi- or non-humanoid avatar designs to enhance the expression of augmented capabilities. & Recommended\\
   \hline
   1.3 & Incorporate simple and universal symbols and metaphors to make augmented capabilities recognizable. & Essential \\
   \hline
   1.4 & Select material and texture that align with augmented capabilities and the avatar’s physical form. & Essential \\
   \hline
   1.5 & Prioritize hands and arms for interaction, while adapting to avatar form and augmented action mechanics. & Essential \\
   \hline
   1.6 & Leverage color semantics to enhance the perception of augmented capabilities. & Recommended \\
   \hline
   1.7 & Use dynamic visual effects to enhance perception and feedback of augmented actions. & Recommended \\
   \hline
   1.8 & Use outfits and accessories to indicate augmented capabilities and triggering interaction methods. & Optional \\
   \hline

       & \textcolor[RGB]{0,176,80}{Physical self-movement} & \\
   \hline
   2.1 & Utilize biomimetic designs or mechanical structures to suggest enhanced physical movement capabilities. & Recommended \\
   \hline
   2.2 & Ensure consistency, operability, and comfort when designing triggering interactions for augmented physical movements based on real-world actions. & Recommended \\
   \hline
       & \textcolor{red}{Bodily transformation} &  \\
       \hline
   3.1 & Identify real-world, cultural, or symbolic references with transformation traits to guide avatar design and make capabilities obvious. & Recommended \\
   \hline
   3.2 & Use materials that align with the nature of bodily transformation to visually reinforce capabilities. & Recommended \\
   
   \hline
       & \textcolor{violet}{Operative action} & \\
       \hline
   4.1 & Utilize key body feature design to highlight operative actions and indicate interaction cues. & Recommended \\
   \hline
   4.2 & Use glowing effects and color changes in interactive body parts to indicate the execution of actions. & Recommended \\
   \hline
       & \textcolor{brown}{Constructive action} & \\
       \hline
   5.1 & Design the avatar’s appearance to establish visual connections with manipulated targets to indicate constructive action capabilities. & Recommended \\
   \hline
   5.2 & Design unique interactions for each aspect of constructive actions to reflect distinct steps in the process. & Recommended \\
   \hline
   
\end{tabular}
}

\label{tab:Guidelines}
\end{table*}
\autoref{tab:Guidelines} presents the proposed avatar design guidelines, organized into five perspectives: general guidelines applicable across all augmented actions and specific guidelines tailored to four categories of augmented actions. 
Each guideline is marked with its level of necessity: 
\textit{Essential} guidelines are fundamental principles that should be prioritized in avatar design. 
\textit{Recommended} guidelines are strongly advised for use in designs. 
\textit{Optional} guidelines can be flexibly applied depending on the specific context.

\subsection{General Guidelines} 
We have summarized 8 general guidelines (G1.1-G1.8 in \autoref{tab:Guidelines}) that are applicable across all augmented actions. These guidelines are organized according to the design process, starting with sources of inspiration and moving on to details regarding materials, colors, and interaction methods.

\textbf{Overall inspiration.} We found that experts drew inspiration from various sources when creating avatars: animals, shared knowledge (such as using red and blue to represent magnetic polarity), famous media (including movies, games, anime, and myths), robotics and technology, unique occupations and cultural references (such as ninja and alchemist), and fantasy creatures (such as dragon and stone monster). 
Drawing inspiration from diverse fields enabled designers to create a wider range of avatar representations for the same augmented action. 
For instance, the ability to fly can be inspired by nature through a biomimetic approach, such as wings shaped like those of swallows, highlighting natural aerodynamics.
Inspiration can also come from science fiction films through technological metaphors like propulsion launchers integrated into shoe soles (A1) (\autoref{fig:formativeoutcomes}~a). 
This observation led to the development of G1.1, emphasizing the need to explore various inspirational sources to enhance the diversity and intuitiveness of avatar designs for augmented actions.

\textbf{Avatar type.} 
We analyzed avatar type, symbols and metaphors, materials and textures, colors, and outfits, leading to the formulation of guidelines G1.2–G1.4, G1.6, and G1.8. 
We categorized the types of avatars and observed that fewer than half were designed as humanoids. 
This suggests that moving beyond human representations more effectively showcases augmented capabilities. 
Experts incorporated animal traits, robotic exoskeletons, and unique body structures and forms to highlight movement, transformation, and environmental interaction, contributing to G1.2. 
For example, a ghost-like floating avatar (A5) could more intuitively convey the capability of intangibility. 
Another common practice is to apply well-known metaphors and symbols to represent augmented actions. 
For instance, ninjas can be applied to symbolize self-cloning, as D7 mentioned in A26 that ``\textit{the ninja is the origin of cloning techniques worldwide}'', illustrating G1.3.

\textbf{Materials, textures, and colors.} 
Experts intentionally selected materials that align with the avatar's augmented capabilities and physical form, reinforcing intuitive visual associations between appearance and capabilities (G1.4). 
For example, metallic and polymer-based materials with fine textures were typically used for mechanical avatars to emphasize strength and durability (A3, A6, A12). 
In contrast, rubber-like or gelatinous textures were applied to avatars with shape-shifting capabilities, highlighting their flexibility and deformation (A5, A7, A8, A9). 
Additionally, the semantics of color were used to enhance the association with specific capabilities. 
For instance, red was used for fire-related abilities (A14, A15), and red and blue represented magnetic poles, contributing to G1.6. 

\textbf{Outfits and accessories.} 
Although outfits and accessories are not mandatory elements in avatar design, we observed that fourteen avatars incorporated these components. 
Among them, eleven used outfits and accessories to communicate avatar capabilities, while three clearly emphasized avatar personalities. 
For example, A20 used workwear and protective gloves to highlight the avatar's identity as an alchemist and a technician, reinforcing its ability for gold transmutation. 
Meanwhile, A1 showcased shoes with buttons and launchers, indicating that pressing the button would activate a propulsion-based flight ability, using the outfit as implicit interaction cues (\autoref{fig:formativeoutcomes}a).
It highlights that outfits and accessories play a crucial role in conveying an avatar's capabilities and suggesting possible interaction methods (G1.8).

\textbf{Interaction.}
As for interaction design, most experts prioritized hands and arms as the primary control points for interactions. 
However, other body parts were also used to improve interaction effectiveness, depending on the avatar's form and the specific augmented actions.
For example, A6 allows users to use a clenched fist to aim and then open the hand to release flying claws, while A15 employs a throwing motion to launch fire-based attacks. 
In some cases, unconventional methods, such as mouth-based interactions in A22 (inhaling to enlarge and exhaling to shrink), aligned more with the balloon toy avatar (see \autoref{fig:formativeoutcomes}~b). Based on these findings, we developed G1.5.

\textbf{Visual effect.} 
Most avatars incorporated visual effects (VFX) for two primary purposes: suggesting inherent capabilities and interactive body parts, and providing visual feedback upon action activation. 
Experts primarily used particle effects to reinforce the nature of the avatar’s powers and interaction methods.
For example, A14 used rising smoke particles to convey intangibility. A16 integrated sparkling particles around hands, guiding users toward hand-based interactions.
In terms of activation feedback, the primary effects were light-based visual cues. For example, glowing coils and wires in the arms indicated generating a magnetic field (A11). Other effects, such as flame bursts, color shifts, and additional particle effects, were also used to confirm that an action has been successfully executed.
These observations led to the formulation of G1.7.

\subsection{Action Type 1: Physical Self-movement}
To effectively communicate augmented physical movement, experts focused on using biomimetic designs and mechanical structures in the avatar’s appearance to convey capabilities (G2.1), and designed interaction methods based on real-world actions to enhance intuitiveness and ease of use (G2.2). 
Two of the six avatars in this category utilized biomimetic features, while three incorporated mechanical structures.  
A2 integrated swallow-wing-shaped arms to suggest the ability to fly, while A4, inspired by crickets, featured insect-like limb angles to emphasize enhanced jumping. 
On the mechanical side, A1, as an example, incorporated a shoe-integrated launcher to indicate propulsion-based flight (\autoref{fig:formativeoutcomes}a). 
In designing triggering interactions, experts often simplified real-world actions to ensure natural consistency while maintaining operability and comfort in VR (G2.2).
For example, A3 simplified jumping triggers to upper body movements, mimicking real-world jump dynamics by having users swing their arms back and lift them forward to activate enhanced jumping. 

\subsection{Action Type 2: Bodily Transformation} 
Experts directly employed real-world, cultural, and symbolic references as design anchors to convey bodily transformation capabilities (G3.1) and aligned them with materials that visually reinforced the transformation traits (G3.2). 
These references served as inspiration and concrete visual metaphors that shaped avatars' appearance and interaction methods. 
For instance, A22 was designed as a balloon toy avatar that utilized the balloon's natural inflation and deflation to demonstrate body size manipulation, paired with a balloon-blowing gesture as the method of interaction (\autoref{fig:formativeoutcomes}b). 

\subsection{Action Type 3: Operative Action}
All experts conveyed augmented operative actions by emphasizing key body features and reinforcing action execution with visual feedback(G4.1, G4.2). 
Distinctive body features, such as large mechanical arms, specialized hand shapes, and textured skin, were incorporated in avatars to suggest operative action capabilities and signal interaction cues(G4.1).  
For example, A13 featured a giant avatar with growth patterns on its arms and legs, visually emphasizing muscle expansion to illustrate its Enhanced Lift capability (\autoref{fig:formativeoutcomes}a). 
As for triggered feedback, experts frequently applied glowing effects and color changes on key interactive body parts to indicate action execution (G4.2). 
A7 and A8 featured color-changing arms to signal activation of extended reach, while A10 and A11 used emissive glowing effects to depict magnetic force generation. 

\subsection{Action Type 4: Constructive Action} 
To express augmented constructive actions, experts designed avatars that visually correlate with the manipulated targets and support multiple interactions to reflect complex action processes (G5.1, G5.2). 
Experts linked avatar appearances with manipulated targets using colors, materials, symbols, outfits, and visual effects. 
For instance, A15 featured a black dragon-inspired humanoid with flaming hair to suggest fire-related abilities (\autoref{fig:formativeoutcomes}d), while A14 incorporated molten lava textures to reinforce the connection to fire-based powers. 
Unlike other action categories, constructive actions often involve multi-stage processes, such as creation, transformation, and destruction. 
Accordingly, each avatar in this category featured at least two distinct interaction methods, each corresponding to a different stage of the constructive process, leading to G5.2. 
For example, A19 was designed with five separate interactions to trigger mountain creation, reshape terrain, destroy mountains, absorb energy, and switch modes. 

%% file: sections/05_Evaluation.tex
\begin{figure*}
    \centering
    \includegraphics[width=1\textwidth]{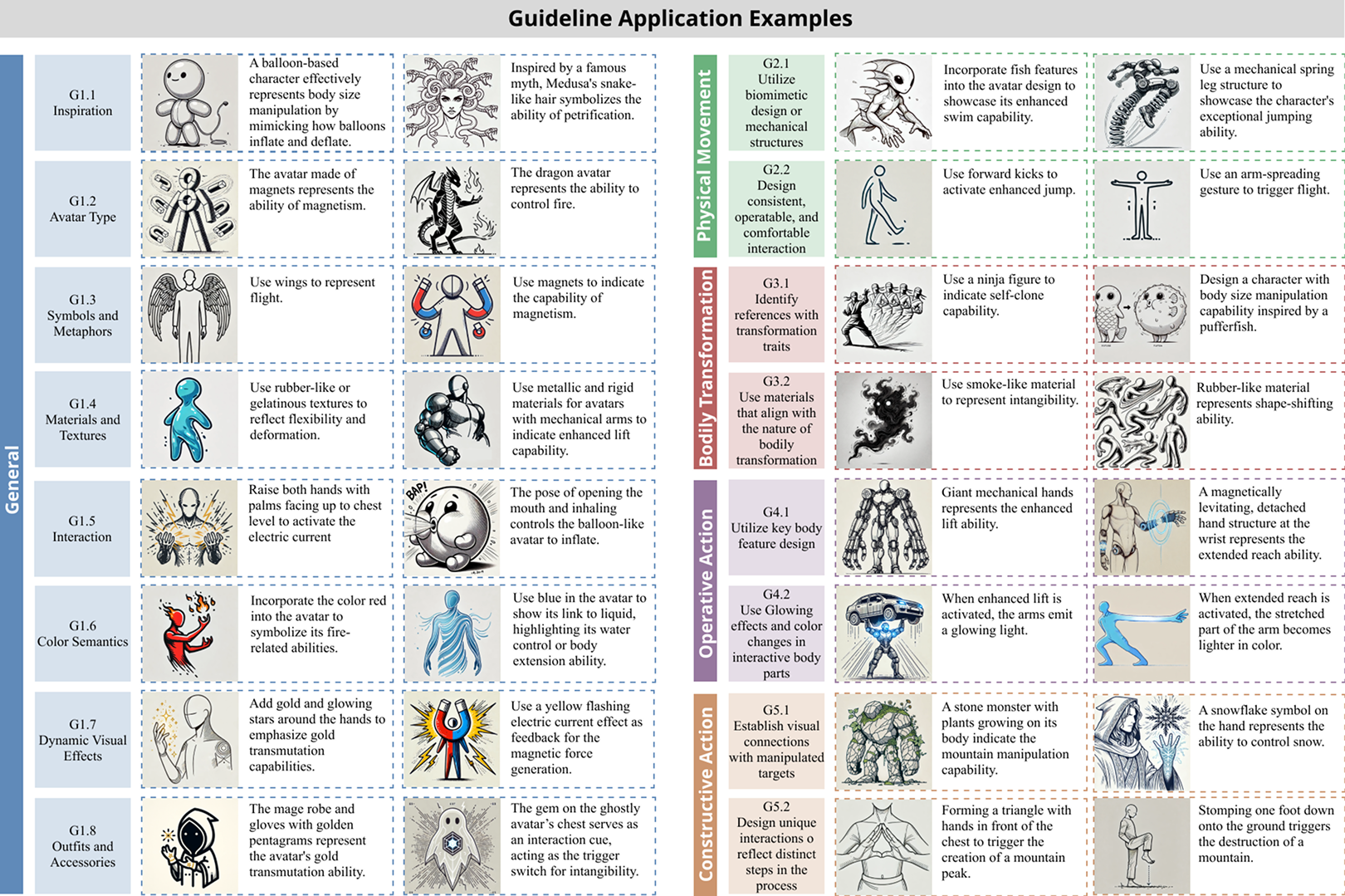}
    \caption{This figure presents the example table of guidelines, where each guideline is accompanied by two illustrative examples. This material was provided to participants in the guideline group to aid their design process.}
    \label{fig:exampletable}
\end{figure*}
\section{Study 2: Design Guidelines Evaluation}
In this section, we evaluated the effectiveness of our design guidelines from two aspects: 
(1) whether avatar designers can understand and apply the guidelines in their design process, and 
(2) whether the guidelines help create avatars that intuitively convey augmented action capabilities and the corresponding interaction methods. 
To address these questions, we conducted a between-subject user study that let participants design avatars for augmented actions with and without guidelines and compared their design outcomes. 
%Next, we demonstrate the study design in detail. 
Throughout this paper, we refer to participants using our guidelines as the \textit{guideline group} and participants without guidelines as the \textit{non-guideline group}.

\subsection{Design Avatars with and without Guidelines}
To evaluate the practical effectiveness of our guidelines for supporting avatar design, we invited novice designers to create avatars for augmented actions with and without access to our guidelines. 
In the following sections, we describe the study setup and provide a detailed analysis of participants’ feedback on their design process using the guidelines.

\subsubsection{Participants. } 
We recruited 24 participants (aged 19 to 30 years, M = 23.25,
SD = 2.75; 6 male, 14 female, 4 undisclosed) through social media advertisements and snowball sampling for this Institutional Review Board (IRB) approved study. 
None of them had participated in Study 1. 
Each participant received \$20 for their participation. 
They had diverse academic backgrounds, spanning fourteen majors, including architectural design, digital media technology, industrial design, computer science, brain and cognitive sciences, etc. 
All participants were highly interested in character design, frequently engaging in character creation as a hobby or through competition experience. 
They demonstrated sufficient hand-drawing skills to sketch avatar concepts independently. 
Additionally, 22 participants had prior experience using VR devices.

\subsubsection{Study Materials} 
All participants received a task description document and two design examples for reference. 
The task description document included an introduction to augmented actions, design tasks, and submission requirements. 
Additionally, participants in the guideline group were provided with supplementary materials, which included a classification table of augmented actions with definitions and example actions for each category, a sheet of our guidelines (\autoref{tab:Guidelines}), and an example table illustrating the application of each guideline (\autoref{fig:exampletable}). 
They were explicitly instructed not to directly replicate designs from the example table, but rather to use it as a reference to inspire their own ideas.

\subsubsection{Task} 
We selected twelve augmented actions, three from each category. 
Four of these actions were retained from the formative study, while eight new actions were introduced to evaluate the applicability of our guidelines. 
Each participant was randomly assigned one augmented action, ensuring that every action was designed by both a participant from the guideline group and a participant from the non-guideline group. 
The instruction given to the participants was to design the visual aspects of avatars, ensuring that their appearance intuitively reflects their augmented action capabilities and interaction methods.
Additionally, participants were tasked with creating interaction methods that strongly link the avatar to its augmented actions.  
The design process took approximately two hours. 
Finally, each participant submitted a design documentation package, which included avatar design sketches, design descriptions, and explanations of the interaction methods. 

\subsubsection{Study Design and Procedure}
We randomly assigned participants into two groups ( non-guideline group: P1-P8 and P21-P24, and guideline group: P9-P20) and conducted the study online. 
Before starting, we introduced the study to each group separately through online meetings conducted in small subgroups of 2 to 4 participants based on individual availability. 
Both groups received the same instructions for the study procedure, augmented action introduction, design tasks, and requirements (about 10 minutes). 
The guideline group received an additional explanation for the augmented action classification, guidelines, and example applications (about 10 minutes). 
Participants received the relevant study materials and were instructed to review them carefully before starting the design task.

Once prepared, participants provided their demographic information and completed the design tasks for an augmented action in two hours independently and without supervision.
They were advised to work in a quiet, distraction-free environment to ensure focus and minimize interruptions. 
After completing the task, participants submitted their results. 
Finally, we collected 25 avatar designs, with each augmented action having two corresponding avatar designs—one created using the guidelines and one without. 
The only exception was the freezing action, for which the non-guideline participant was enthusiastic about avatar design and created two avatars instead of one. 
We carefully reviewed all avatar designs and found no instances of participants directly replicating designs from the example table. All avatar designs are included in the supplementary materials.

After participants submitted their results, we collected feedback from both groups. 
For the guideline group, we first asked participants to rate the clarity (\textit{To what extent is the guideline clear and easy to understand?}) and usefulness (\textit{To what extent is the guideline useful and helpful in your Avatar design process?}) of the guidelines. 
Each participant evaluated all eight general guidelines and the two category-specific guidelines corresponding to their assigned augmented action. 
Then, we interviewed participants about their use of the guidelines, how the guidelines impacted their design process, and any suggestions for improvement (about 10 minutes). 
For the non-guideline group, we interviewed participants about their challenges in the design process. 
We then introduced them to the augmented action classification, guidelines, and example applications and asked them how having access to these guidelines might have improved their designs. 
It took about 20 minutes. 
Finally, we asked them to rate the clarity and usefulness of the guidelines retrospectively (\textit{To what extent do you think this guideline would have been helpful in your design process?}), evaluating the same eight general guidelines and the two category-specific guidelines relevant to their assigned augmented actions.

\subsubsection{Guideline Usage}
To understand the differences in avatars designed by guideline and non-guideline groups, we counted which guidelines were reflected in the avatar designs created by the two groups. 
Three authors independently coded the avatar designs according to whether they satisfied each guideline and then cross-checked their codes and resolved disagreements through discussion.  
Inter-rater reliability was assessed using Fleiss's kappa, resulting in a score of 0.49, which indicates moderate agreement. 
The results showed that avatar designs from the guideline group incorporated an average of 8.92 guidelines (SD = 0.79) per design, while those from the non-guideline group included an average of 5.85 guidelines (SD = 3.00).
As the non-guideline group did not have access to the guidelines, it was expected that fewer guidelines would be satisfied in their designs.
However, from the comparison, we found that G1.2 and G1.8 were the guidelines that were most easily overlooked by the non-guideline group, indicating that participants typically did not consider designing non-humanoid avatars and often did not use outfits or accessories to hint at capabilities and interaction methods. 

Additionally, other guidelines on detailed design aspects, such as materials, colors, and visual effects, as well as category-specific guidelines, were often missing in the non-guideline group. 
This comparison also confirms that the guidelines helped participants incorporate more relevant and coherent design elements. 
When structured guidance was available, participants were more likely to consider aspects that might not usually come to mind during the design process and were less likely to miss important design elements. 
\begin{figure}
    \centering
    \includegraphics[width=0.9\linewidth]{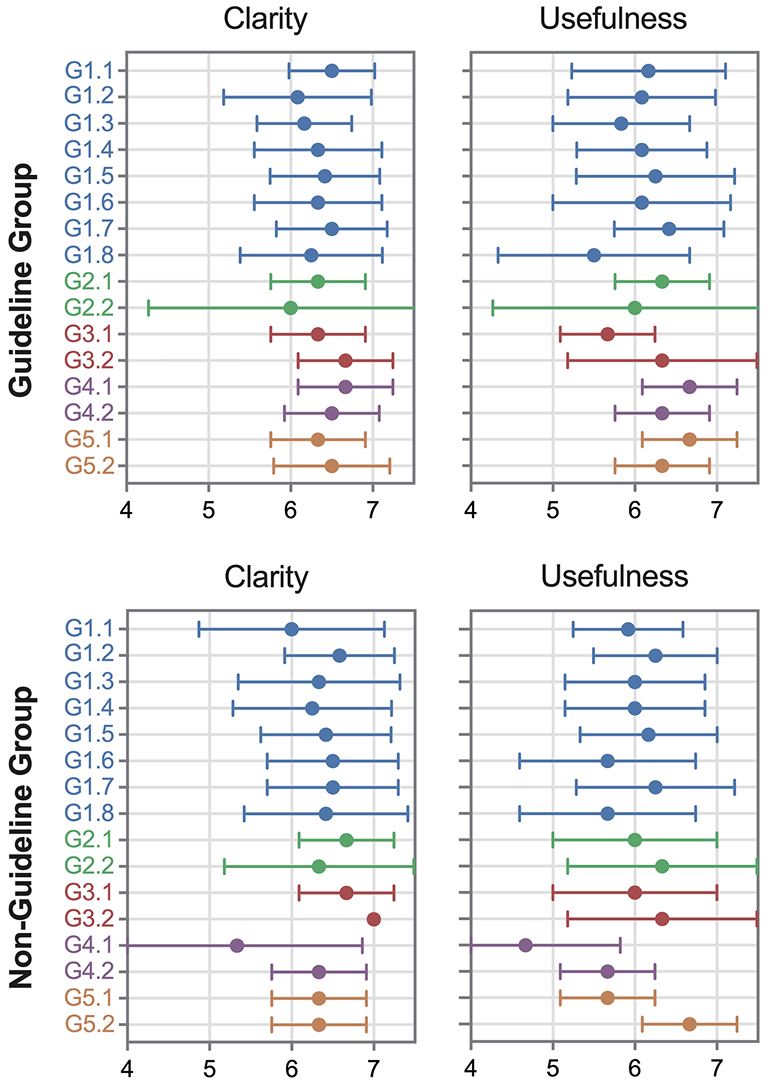}
    \caption{Guideline group and non-guideline group ratings with standard deviation on guidelines' clarity and usefulness.}
    \label{fig:guidelinerating}
\end{figure}
\subsubsection{User Feedback on the Guidelines} 
Generally, all participants in both the guideline and non-guideline groups rated the clarity and usefulness of the guidelines highly, with almost all guidelines receiving an average score of 5 or above on a seven-point Likert scale (\autoref{fig:guidelinerating}). 
In the post-study interviews with the guideline group, we identified the specific guidelines each participant used in their avatar designs. 
We summarized their strategies for applying these guidelines, how the guidelines supported their design process, and their suggestions to improve the guidelines. 

\textbf{A high rate of guideline adoption by participants suggests their practical relevance and ease of use.} 
Each participant used at least eight guidelines in their avatar designs. 
Three participants adhered to all ten relevant ones, including eight general guidelines and two specific to their assigned action category. 
It suggests that the guidelines were easy to understand, follow, and apply. 
Among the guidelines, G1.2 and G1.8 were the least utilized. Three participants did not adopt G1.2, with P14 stating: ``\textit{I thought about using a semi-humanoid form but couldn't come up with one.}'' 
It suggests that while G1.2 successfully encouraged participants to explore diverse avatar forms, some found it challenging to apply effectively in their designs. 
As for G1.8, which was marked as optional, most participants chose to apply it, with only three opting not to use it. 

\textbf{The guidelines enhanced creativity during the ideation process.} 
Several participants (P9, P12, P14, P15, P17, P20) emphasized how the guidelines enhanced creativity by broadening their perspectives and opening up new design possibilities. 
P9 noted, ``\textit{the guidelines help me go deeper and broader in my thinking.}'' 
Similarly, P20 remarked that the guidelines ``\textit{brought a lot of inspiration and helped me create a more expressive and cohesive avatar design.}''
P12 specifically praised G1.1 for encouraging her to draw inspiration from everyday life and nature, which broadened her creative possibilities.
P14 described the guidelines as providing “heuristic prompts” that opened up possibilities he might not have considered otherwise, making the ideation process faster. 
Likewise, P15 treated the guidelines as a valuable reference: ``\textit{They showed me possible ways to express a capability, which expanded my imagination.}'' 
She further noted that the guidelines were particularly helpful for novice designers, promoting thought from multiple dimensions and perspectives.
 
\textbf{The guidelines streamlined the design process by clarifying design goals and supporting detail refinement.}
P10, P13, P16, P18, and P19 indicated that the guidelines clarified which aspects needed to be addressed, offering clear direction from the outset. 
P10 considered the guidelines ``\textit{a design paradigm listing aspects that need to be addressed to create an excellent avatar.}'' 
Echoing this view, P16 reflected that ``\textit{without the guidelines, I wouldn't know what exactly to design. I might go through many materials and ideas, but still miss the point.}'' 
Additionally, several participants (P10, P11, P12) highlighted that the guidelines helped them transform vague ideas into concrete designs by emphasizing specific design elements such as color, material, and visual effects. 
P12 mentioned that these specific guidelines reminded her to consider finer design details, such as how color and lighting could be used to convey capability. 
P10 specifically praised G1.7 for encouraging the use of visual effects to make an avatar’s capabilities and interactive body parts immediately recognizable at a glance. 
P11 described the guidelines as a clear organizational framework: ``\textit{It felt like having an assignment outline. I had the concept in mind, and the guidelines helped me fill in the details.}''

Additionally, we observed three main utilization strategies: \textit{Pre-reading with post-design validation}, \textit{Active reference during design}, and \textit{hybrid use}. 
Two participants (P9, P12) reviewed the guidelines before starting to design. They then completed their designs without consulting the guidelines, but revisited and referred to them afterward to validate and refine their designs. 
In contrast, most participants (P10, P11, P13, P14, P16, P18, P19) actively referenced the guidelines throughout the design process.  
P15, P17, and P20 utilized a hybrid approach by reviewing all guidelines to determine a design direction and referring to specific sections for additional guidance or validation during the design. 

\textbf{Potential improvement: to provide guideline materials that vary in detail to accommodate designers with different levels of expertise.} 
Several participants noted that our guideline materials did not provide enough detail for individuals without a background in avatar design. 
They expressed a need for more concrete examples and visual references to apply the guidelines effectively. 
For instance, P9 noted that, as a non-professional designer, she had to rely heavily on external resources to supplement the textual guidance: ``\textit{Even with the example table, I couldn't visualize my idea into sketches without additional references.}'' 
In contrast, some experienced participants felt that the application examples constrained their creativity. 
P14 mentioned that he seldom referred to the example table during design: ``\textit{The more I looked at the examples, the more I felt my thinking became boxed in.}'' 
These observations indicate that designers at different professional levels may benefit from varying detail levels of guidance. 
One potential solution is to offer tiered materials through an interactive format, such as a website, where users can explore different levels of explanation and examples based on their needs. 

\subsection{Avatar Intuitiveness Evaluation through an Online Survey}
To compare the avatars designed with and without our guidelines, we invited 48 general audiences (24 males, 24 females) to evaluate the avatars' capability and interaction intuitiveness through an online survey.
We recruited them through online advertisements and snowball sampling, ensuring none were involved in designing avatars. 
Participants were evenly divided into two groups, each with an equal gender distribution (12 male, 12 female). 
One group evaluated 17 avatars representing eight augmented actions spanning all four capability categories, while the other group evaluated 8 avatars representing four augmented actions, also covering all four categories. 
This grouping strategy ensured full category coverage for participants while keeping the survey duration manageable. 
Evaluating all 25 avatars would take over an hour, so splitting participants into two survey lengths helped reduce fatigue without sacrificing the diversity of capabilities assessed.
In both groups, each action was represented by two avatar designs, one designed with guidelines and one designed without guidelines, except for the freezing action, which included two avatars designed without guidelines. 

\subsubsection{Survey Design and Evaluation Procedure}
The evaluation survey included three types of questions: 
(1) short-answer questions for capability inference, 
(2) Likert-scale ratings assessing the intuitiveness of each avatar's capability and interaction, and 
(3) pairwise comparative selections between avatars designed with and without the guidelines. 
The full set of survey questions is provided in the supplementary materials.

Initially, participants were presented with an avatar design image with only material annotations. 
They were asked to infer the augmented capability based on the image by answering the open-ended question: \textit{Based on the avatar's design and description, what augmented capabilities (or superpowers) do you think it possesses?} 
After submitting their response, they were shown the correct capability with corresponding interaction methods and asked to rate how intuitively the avatar's design conveyed this capability and interaction methods. 
The presentation order of avatars was counterbalanced, with half of the participants evaluating avatars designed with guidelines first and the other half evaluating avatars designed without guidelines first. 
Participants were blind to the design condition of each avatar to avoid bias in their evaluations. 
We adopted this counterbalancing method to ensure that avatars reflecting the same augmented action were evenly spaced apart, minimizing sequence effects and maintaining fairness across participants. 
In contrast, a fully randomized order could place two avatars designed for the same action consecutively, making it easier to infer the intended capability in the second trial. 
Following the rating section, participants proceeded to the comparative selection section, where they directly compared avatars designed for the same augmented action. 
Participants were presented with side-by-side images of avatars created for the same action and their respective interaction methods. 
They were then asked to answer two comparative questions: 
(1) \textit{Which avatar design more intuitively reflects the capability? }
(2) \textit{Which avatar design more intuitively reflects the interaction method?} 

\subsubsection{Comparison of Intuitiveness Evaluation Results}
\begin{figure}
    \centering
    \includegraphics[width=0.9\linewidth]{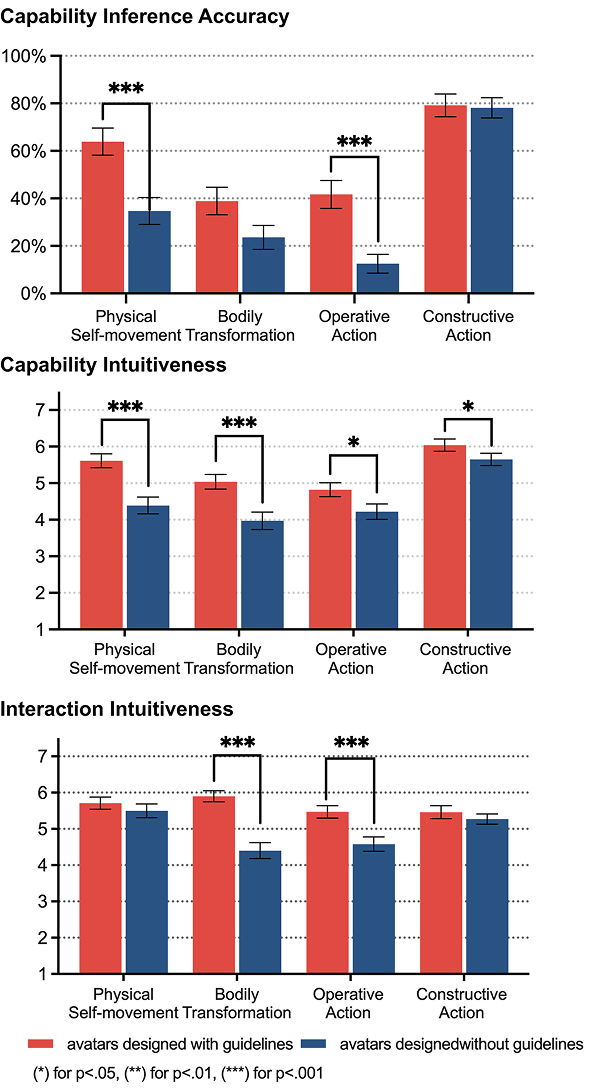}
    \caption{General audiences’ evaluation results on the intuitiveness of avatars designed with and without guidelines across four augmented action categories. The top chart shows the average capability inference accuracy with standard errors. The middle chart presents the average capability intuitiveness ratings, and the bottom chart shows the average interaction intuitiveness ratings, both with standard errors.}
    \label{Evaluationresults}
\end{figure}
Participants were asked to complete the survey on a computer to ensure optimal visibility of the images for accurate evaluation. 
Before starting, they were given a response example and instructed to review it carefully. 
Participants who evaluated 17 avatars answered 59 questions in total, which included both individual avatar ratings and comparative selections. This survey took approximately 45 minutes to complete. 
Participants who evaluated 8 avatars answered 28 questions, and the survey took approximately 25 minutes to complete. 

We evaluated avatar intuitiveness through three metrics: capability inference accuracy, capability intuitiveness ratings, and interaction intuitiveness ratings. 
Additionally, we reported the results in the comparative selection section, where participants directly compared avatars designed with and without the guidelines regarding capability and interaction intuitiveness. 
To calculate the accuracy of participants' inferences, we coded responses to the capability inference open-ended question as either correct or incorrect. 
Three authors independently coded the participant responses, then compared the coding results and discussed to reach a consensus, ensuring reliability and consistency. 
\autoref{Evaluationresults} presents the means and standard errors for these metrics of avatars designed with and without guidelines for four augmented action categories. 
Overall, participants perceived that avatars designed with guidelines were more intuitive in suggesting augmented action capabilities and interaction methods than those designed without guidelines.
This was demonstrated by higher capability inference accuracy, intuitiveness scores of capability and interaction, and a greater number of selections as the more intuitive design in the comparative selection section. 
To analyze the differences, we used Wilcoxon signed-rank tests for paired comparisons in the Physical Movement, Bodily Transformation, and Operative Action categories and the Mann–Whitney U test for unpaired comparisons in the Constructive Action.
This choice of non-parametric tests was motivated by the non-normal distribution of the data. 
In the Constructive Action category, an unpaired comparison was required because one non-guideline participant enthusiastically created two avatars, resulting in an unequal number of avatar designs between the two groups.

\begin{figure*}
    \centering
    \includegraphics[width=1\textwidth]{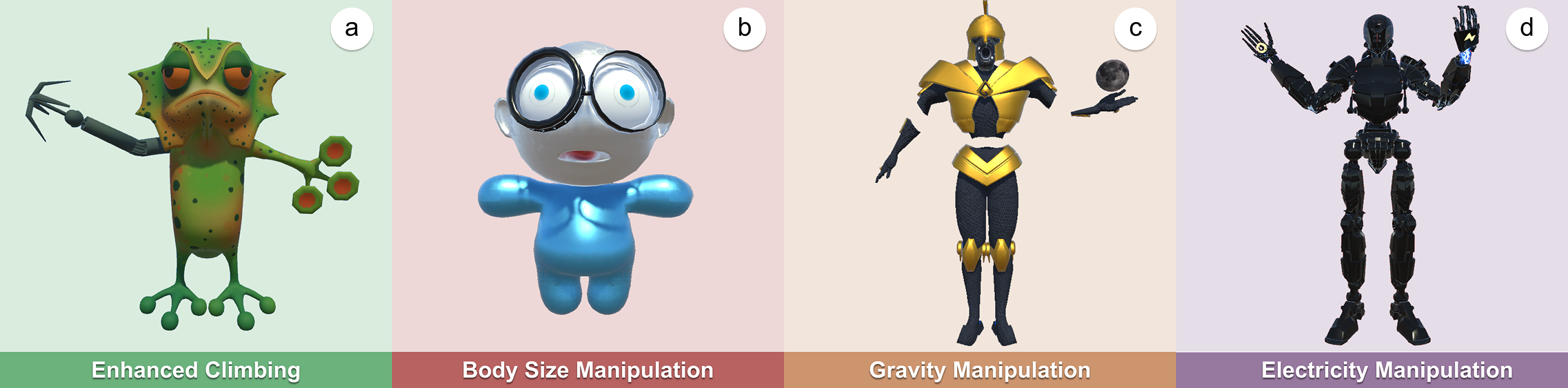}
    \caption{Four avatars designed with our guidelines, each representing a distinct category of augmented action: 
    (a) a lizard-like avatar with a mechanical hook to suggest enhanced climbing; 
    (b) a balloon-like avatar with exaggerated proportions and oversized glasses to suggest body size manipulation; 
    (c) a faceless humanoid with floating body parts and a levitating planet to suggest gravity manipulation;
    (d) a robotic avatar with lightning bolt symbols on the back of each hand to suggest electricity manipulation.
    }
    \label{Avatar}
\end{figure*}
In terms of capability inference accuracy, avatars designed with guidelines achieved higher average accuracy across all categories than those without guidelines. 
The differences were statistically significant for avatars in physical movement ($p < .001$) and operative action ($p < .001$) categories, and marginally significant for bodily transformation ($p = .072$). 
No significant difference was found for constructive action, which may be attributed to the familiarity with such capabilities in popular media. 
Capabilities such as freezing, fire manipulation, and desert manipulation are commonly depicted in films and video games. 
As a result, participants in both groups may have relied on similar reference imagery, which could reduce differences in inference accuracy. 
Nevertheless, avatars designed with guidelines were still perceived as more intuitive. 
Participants gave significantly higher capability intuitiveness ratings to guideline-based avatars in all four categories: physical movement ($p < .001$), bodily transformation ($p < .001$), operative action ($p < .05$), and constructive action ($p < .05$).
Together, these findings confirm that our guidelines effectively assist users in designing intuitive avatars, enabling augmented action capabilities to be recognized directly from the avatar’s appearance in most cases.

As for interaction intuitiveness, the avatars designed with guidelines in the bodily transformation and operative action categories received significantly higher ratings ($p < .001$) than those without. 
However, no significant difference was found in the physical movement and constructive action categories. 
This phenomenon may be attributed to the fact that it is easier to find references in real-world experiences or popular media when designing interaction methods for physical movement and constructive action (e.g., enhanced jumping and fire manipulation). 
As a result, participants could intuitively associate these actions with their interaction methods, even without explicit visual cues. 
In contrast, bodily transformation (e.g., body size manipulation) and operative action (e.g., gravity manipulation) involve less familiar or more abstract mechanisms, making conveying interaction methods solely through avatar appearance more challenging. 
These findings emphasize the significance of our guidelines in assisting the design of avatars to suggest interaction methods for abstract or unfamiliar augmented actions. 

Regarding comparative selection results, we summarized participants’ pairwise choices between avatars designed with and without guidelines. For each of the eight augmented actions, 24 participants selected one avatar for better capability intuitiveness and one for better interaction intuitiveness, resulting in 288 selections for capability and 288 for interaction. 
For capability intuitiveness, avatars designed with guidelines were selected 165 times, while avatars designed without guidelines were selected 79 times and 44 times for the same. 
For interaction intuitiveness, avatars designed with guidelines were selected 146 times, while avatars designed without guidelines were selected 103 times, with 39 times for the same. 
These results further validate the effectiveness of our guidelines in assisting designers to design avatars that intuitively communicate both augmented action capabilities and interaction methods. 

%% file: sections/06_Implementation.tex
\section{Towards Practical Applications of Intuitive Avatars}
To verify the practical applicability of our guidelines, we implemented four avatars designed with our guidelines in VR, each illustrating a distinct augmented action. 
Two avatars were adapted from high-performing designs in the guideline group of Study 2, and the other two were designed by the authors following the guidelines to demonstrate the specific augmented actions. 
The avatars were not designed for specific application contexts; instead, we selected four example VR scenarios, including sports games, navigation, and educational scenarios, to illustrate how these avatars could be used in different settings. 
As such, these scenarios serve as demonstrations rather than optimized pairings, and alternative avatar designs may be more suitable for certain applications. 
We then conducted a demonstration study to gather user feedback while using these avatars in four scenarios.

\subsection{Avatar Implementation}

The four avatars we implement in VR were designed for augmented actions from four categories. 
Two avatars were adapted from high-performing designs in the guideline group of Study 2: a body size manipulation avatar from the bodily transformation category and a gravity manipulation avatar from the operative action category. 
The other two avatars were designed by the authors following our proposed design guidelines: 
an enhanced climbing avatar from the physical movement category, and an electricity manipulation avatar from the constructive action category. 
All avatars were 3D-modeled and implemented based on their sketches and descriptions. 
We used Meta XR Movements SDK to realize full-body tracking and face tracking, and the Interaction SDK to enable gesture recognition.

\subsubsection{The Enhanced Climbing Avatar}
This avatar takes the form of a lizard with asymmetrical hands~(\autoref{Avatar}a): 
the left hand consists of three suction-cup-like fingers, while the right hand features a telescopic grappling hook. 
Users aim by facing their right palm toward a target and launch the hook by performing a forward-throwing fist gesture, mimicking the motion of throwing the grappling hook outward.
Once attached to a climbable surface, the avatar’s body is pulled toward the target. 
Upon arrival, users can simulate climbing by stepping in place and performing alternating hand-over-hand motions. 

\subsubsection{The Body Size Manipulation Avatar}
This avatar resembles an inflatable balloon toy with exaggerated proportions and a glossy, rubber-like surface to suggest scale adjustment~(\autoref{Avatar}b). 
It wears oversized glasses, with the right lens mimicking a camera zoom lens to imply visual scale adjustment. 
Users can control their body size with gestures inspired by balloon blowing: 
opening the mouth wide to simulate a deep inhale causes the avatar to expand and become more transparent, while exhaling shrinks the body and reduces its transparency. 
To realize this functionality, we tracked users’ mouth movements using the Meta Quest Pro in combination with the Movement SDK.

\subsubsection{The Gravity Manipulation Avatar}
This Avatar features a faceless humanoid design with floating body parts~(\autoref{Avatar}c). 
Its arms and torso appear disconnected and suspended in midair. 
The avatar wears a golden helmet, within which swirls a black hole. 
A small planet hovers above the avatar’s left hand and moves in sync. 
Users can control gravity by adjusting the height of the left hand: raising the hand decreases the gravitational force, while lowering it increases it.

\subsubsection{The Electricity Manipulation Avatar}
This avatar takes the form of a humanoid robot~(\autoref{Avatar}d). 
A lightning bolt symbol on the back of each hand and glowing circular patterns on the palms highlight its electrical abilities. 
Users can generate electricity by bringing their palms together, which activates a flickering electric effect between their hands to signal activation. 

\subsection{Demonstrative Application Scenarios}
\subsubsection{Encouraging Body Engagement in a VR Climbing Game}
Inspired by prior work on human augmentation in sports~\cite{7891106} and the impact of avatar appearance on physical performance in VR~\cite{10.1145/3410404.3414261}, we implemented the enhanced climbing avatar in a VR rock-climbing game to promote full-body engagement. 
While traditional games often require onboarding tutorials, this mechanical lizard-inspired avatar, paired with full-body tracking, allows users to understand the gameplay mechanics intuitively without explicit instruction. 
Players can use the avatar’s grappling hook to swing across distant cliffs or climb directly along rocky surfaces. 
As shown in~\autoref{fig:teaser}a, the user actively engaged the entire body through step-in-place and hand-over-hand motions to climb up. 
The avatar creates a playful and novel experience while encouraging users to exercise more during gameplay.

\subsubsection{Adjusting Navigation Speed in VR}
Inspired by prior work on Ground-Level Scaling in VR~\cite{abtahi2019m}, we implemented the body size manipulation avatar in a VR navigation scenario to support intuitive modulation of movement speed. 
Leveraging full-body tracking, users can physically walk in the real world to navigate the virtual environment. 
They control their movement speed by adjusting the size of their avatar. Enlarging the avatar makes the environment seem smaller, allowing users to cover more ground with each step and offering a wider field of view. 
In~\autoref{fig:teaser}b, the user is inhaling with a wide open mouth to enlarge the body. 
On the other hand, reducing the avatar's size decreases movement speed, facilitating more precise navigation and closer inspection of environmental details. 
The balloon-like avatar form and the blowing-inspired interaction create a strong conceptual link between appearance, capability, and interaction methods. 
This design can facilitate easy recall, minimize confusion with other input modalities, and enhance users’ overall sense of control and embodiment during navigation.

\subsubsection{Enhancing Comprehension of Gravity in VR learning}
Inspired by prior work emphasizing how self-avatars can aid in understanding abstract concepts~\cite{parmar2022immersion}, we implemented the gravity manipulation avatar in a VR educational scenario to enhance understanding of gravitational concepts. 
In this scene, users observe a toy car navigating the surface of a miniature Earth. 
By raising or lowering the floating planet in the avatar’s left hand, users can adjust the simulated gravitational force in real-time and see how it directly affects the car’s motion. 
In~\autoref{fig:teaser}c, the user reduces the gravitational force, causing the toy car to lift off the ground. 
This cosmic-inspired avatar, with gravity manipulation capability, can make the learning process more memorable and engaging, leading to better conceptual understanding and retention.

\subsubsection{Enhancing Understanding of Circuit Connectivity in VR Learning}
We implemented an electricity manipulation avatar in a circuit learning scenario to enhance the learning experience, making it more enjoyable and safer. 
In the scenario, there is a simple circuit with two light bulbs connected in series, featuring two interruption points. 
To interact, users can generate electricity by bringing their palms close together, triggering a visual effect of electricity on their hands. 
Then, they can connect the circuit by touching the broken segments (\autoref{fig:teaser}d), experimenting with different configurations to manipulate the current flow, and observing how it affects bulb brightness in real-time. 
The robot-like appearance and vibrant visual feedback can instantly capture users' attention. 
Connecting virtual circuits with their hands is simple for users to understand and removes safety concerns associated with real-world circuit experimentation. 
This avatar can foster active participation, promote immersive exploration, and make the learning experience enjoyable.

\subsection{Demonstration Study} 
To further examine how intuitive avatars designed with our guidelines perform across various VR scenarios, we conducted a demonstration study on four applications.
\subsubsection{Participants}
We recruited 12 new participants (aged 19 to 25, M = 23.23, SD = 1.96; 6 males, 6 females) who did not attend previous studies on campus. 
Each participant received a compensation of \$15 for their participation.
All participants had prior experience using VR headsets.

\subsubsection{Procedure}
Before starting, we introduced the study to each participant and collected their demographic information. 
Once prepared, participants entered a virtual environment featuring a mirror, where they could observe their embodied avatar in full view. 
In this mirror scene, participants were first asked to identify the avatar's augmented action capabilities from a list of eight options displayed in VR. 
We applied this multiple-choice format because correctly identifying the intended capability from multiple alternatives still requires meaningful recognition, even though it is easier than an open-ended response. 
Our earlier online survey, which used open-ended guessing, already demonstrated that the guidelines effectively support intuitive interpretation. 
In the VR demonstration, our goal was therefore to verify whether participants could quickly identify the correct capability and to assess how these avatars function in various VR scenarios. 
This format also kept the study duration manageable and prevented participants from spending excessive time on open-ended guessing. 
Participants responded verbally. 
If their answers were incorrect, the experimenter would promptly reveal the correct response.
Participants were then asked how to trigger the capability, verbally describing and demonstrating the physical gesture. 
Correct answers were recognized; otherwise, the correct interaction methods were explained. 
We recorded whether each participant correctly inferred the capability and interaction method.
Participants then engaged with the specific application scenario linked to the avatar.
They were given time to explore and interact with the environment freely. After each experience, participants had a short break, during which we conducted a short interview, focusing on their experience in the specific applications. 
This process was repeated for all four avatars and their associated application scenarios. 
At the end of the session, participants participated in a final semi-structured interview to reflect on their experiences with the four applications. 
The whole study lasted 30-40 minutes.

\subsubsection{Apparatus}
We implemented the applications with a Quest 3 headset in Unity 6000, powered by an Intel Core i5 CPU and an NVIDIA GeForce RTX 3060 GPU. 
We integrated Meta XR Movement SDK and Interaction SDK to support full-body tracking and interaction features. 
We used a Wizard-of-Oz approach for facial tracking and foot tracking in the study and focused on the avatars' appearance and interaction design. 
The study was conducted in an indoor lab office setting. 
Participants stood and wore the VR headset to complete the study. 
A physical space of approximately $2\,\mathrm{m} \times 5\,\mathrm{m}$ was provided, allowing participants to move freely and perform spatial interactions during the VR experience. 

\subsection{Results}
Overall, participants successfully inferred the avatars’ capabilities and interaction methods, suggesting that guideline-based avatar designs were highly intuitive in VR implementation. 
Across all four applications, participants also provided consistently positive feedback, reinforcing the practical value of intuitive avatars in diverse VR contexts.

Most participants accurately identified each avatar’s augmented capability (44 out of 48 trials) and interaction method with minimal explanation (42 out of 48 trials). 
Four participants were initially unsure about the body size manipulation avatar, but after clarification, quickly made sense of the capability and correctly inferred the interaction. 
For the gravity manipulation avatar, four participants did not precisely describe the intended interaction but correctly focused on the floating planet as an interactive element. 
Only one participant incorrectly identified the interaction method for the electricity manipulation avatar, indicating that the palm-based indicator could be made more noticeable. 
All participants correctly inferred all other capabilities and interactions. 
These results show that guideline-based avatars effectively communicate augmented actions when implemented in VR. 
Even first-time users found them easy to understand and engage with naturally, without explicit instructions.

In post-study interviews, all participants agreed that the intuitive avatars made it easy to infer the capabilities and the corresponding interactions from appearance. 
P9 appreciated that ``\textit{they eliminate the need for a tutorial and lower the entry barrier.}'' 
In the climbing game, participants noted that the avatar made their experience more engaging and enjoyable. 
The grappling hook was particularly praised for being visually striking and functional. 
P7, P10, and P12 also emphasized that the avatar encouraged full-body exercise and made them feel more physically active. 
For the navigation scenario, most participants found that balloon-like avatar with a breathing-based mouth interaction to be ``\textit{creative and controllable,}''  providing an immersive and effective way to modulate navigation speed.  
P7 highlighted that using breathing as an input was non-intrusive and did not disrupt other interactions. 
In educational scenarios, participants generally agreed that intuitive avatars enhanced engagement and made abstract concepts more accessible. 
Regarding the gravity learning scenario, P5 and P7 noted that through visual metaphors like floating planets and responsive car trajectories, the avatar helped users visualize the effects of gravitational change. 
Several participants also felt the cosmic-themed design would appeal to younger learners (P6, P7, P12). 
However, some participants (P4, P11) felt the avatar's role was limited. While hand gestures were clear and effective, the avatar itself did not significantly contribute to the learning process. 
In the electricity learning scenario, several participants (P9, P10, P12) appreciated the avatar’s strong visual effects and “hands-on” feel, making circuit exploration safer and more engaging. 
However, others (P3, P11) questioned whether the avatar clarified core concepts such as current direction or voltage logic and suggested the need for more instructional elements like polarity indicators. 

These findings demonstrate that intuitive avatars designed with our guidelines can enrich user experience across diverse VR applications. 
While the avatars enhanced engagement, immersion, and joy in interactive and playful contexts, their effectiveness in educational scenarios depended more on alignment with instructional content and learning goals.

%% file: sections/07_Discussion.tex
\begin{figure*}
    \centering
    \includegraphics[width=0.8\textwidth]{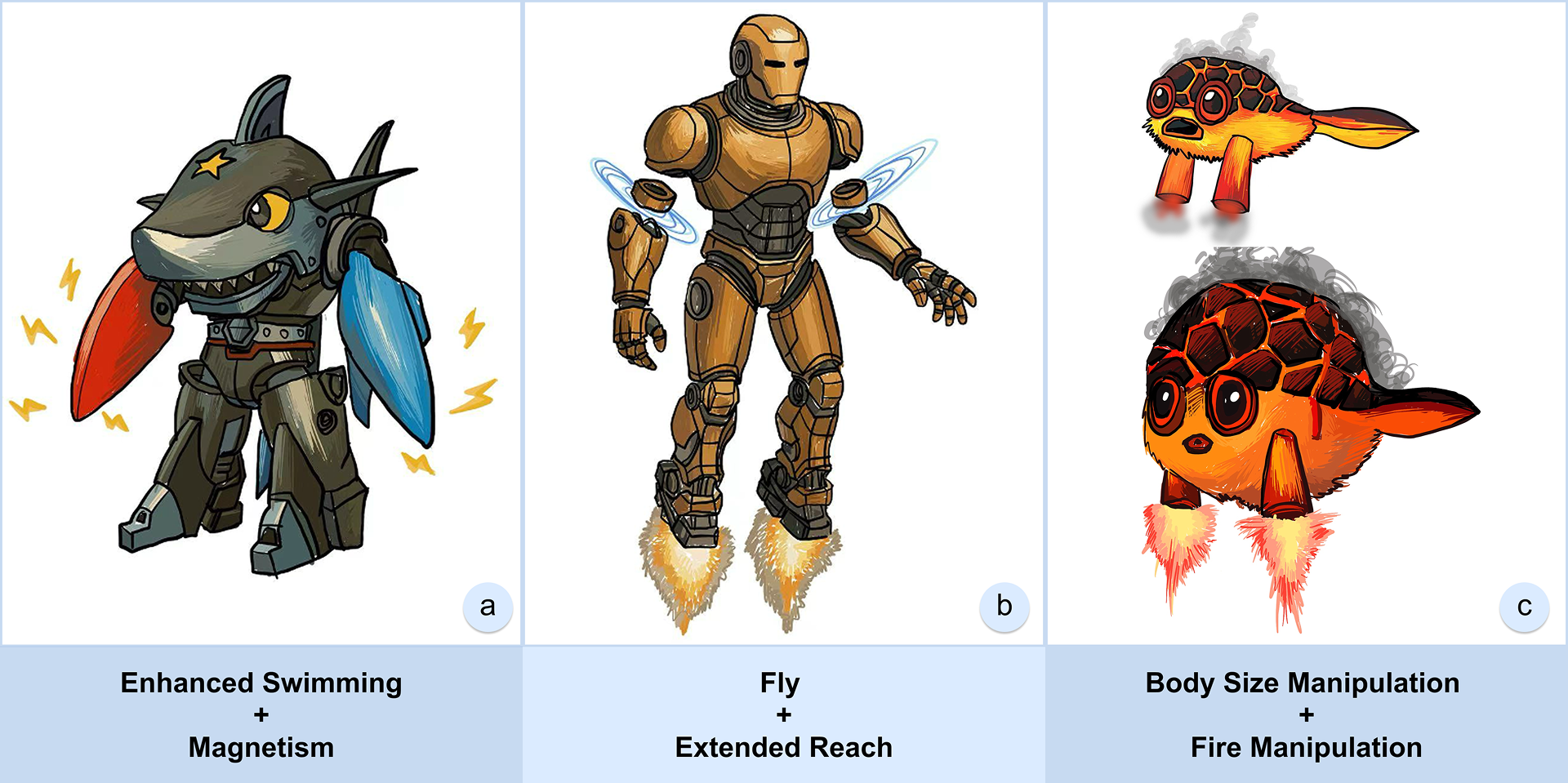}
    \caption{Multi-capability avatars designed following our guidelines. (a) A mechanical shark avatar featuring a metallic texture and red-blue fins, suggesting magnetism and enhanced swimming abilities. 
(b) A robot avatar with floating arms and foot thrusters indicating extended reach and flight capabilities. 
(c) A red pufferfish avatar with lava textures and black smoke, representing body size manipulation and fire control.
    }
    \label{Multicap}
\end{figure*}
\section{Discussion}
Our study developed and validated a set of design guidelines for creating intuitive avatars that communicate augmented capabilities and their corresponding interaction methods. 
Through expert-driven design tasks, a controlled evaluation study, and real-world VR applications, we formulated these guidelines and demonstrated their effectiveness and applicability in various scenarios. 
In this section, we reflect on key themes arising from our findings, ranging from individual differences in guideline usage and expectations to opportunities for improving guidelines and delivery to their broader applicability beyond VR. 
We also discuss the limitations of our current work and directions for future research.

\subsection{Personalized Guidelines for Individual Preferences} 
\textbf{Our findings showed significant individual variations in the use and expectations of the guidelines, highlighting the necessity for more personalized delivery to accommodate diverse workflows and needs.}
We observed that participants used the same set of guidelines in different ways. 
Some designers initially reviewed the guidelines to frame their ideas and later used them for validation, while others referred to them continuously or followed them sequentially as a design outline. 
Additionally, participants expressed conflicting preferences regarding guideline details and examples. 
Novice designers appreciated concrete examples and explanatory language and even expressed the need for more examples. 
In contrast, experienced designers found the examples too specific, which could limit creative exploration. 
These observations reflect clear individual differences in how designers engage with and benefit from the guidelines. 
Therefore, we consider it as future work to make guidelines adaptive to users’ habits and expertise levels or to allow users to personalize them on demand. 
In this manner, we expect the guidelines to better support diverse design workflows and preferences, ultimately streamlining the overall process.

\subsection{Intended User Groups of the Guidelines}
Our guidelines are designed for both novice designers and professional designers. 
\textbf{For novice designers, the guidelines provide clear, step-by-step instructions from finding inspiration to detailed refinement, including color, materials, and visual effects.}  
Following the guidelines and the example table, novices can more easily and effectively express an avatar's augmented actions through various design elements.
This helps reduce uncertainty in early design stages and smooths the design process.
\textbf{For professional designers, the guidelines work more like a checklist.}
They may not need to follow every guideline while designing, but the guidelines can help them review and refine their work afterwards, ensuring that all relevant design considerations are included and applied appropriately.
Our guidelines can also facilitate effective communication within design teams. 
E4 from Study 1 noted, ``\textit{With the guidelines, I can tell my colleagues why a design feels unclear or doesn't work well, which is usually hard to explain without a shared set of terms.}'' 
Overall, the guidelines support designers with different levels of proficiency. They provide novices with concrete guidance during creation, assist professionals in reviewing and adjusting their designs, and offer a shared language that facilitates discussion and collaboration within teams.

\subsection{Application-Specific Considerations in Applying the Guidelines}
While our guidelines successfully supported the design of intuitive avatars across a range of augmented actions, applying them in specific application contexts revealed further complexities. 
In Study 3, participants generally found that the avatars effectively conveyed their capabilities and interaction methods. 
However, some participants noted that the avatars could better align with the goals or constraints of particular scenarios. 
For example, in educational applications, several participants felt that the avatars did not significantly contribute to conceptual understanding. 
Some suggested incorporating instructional elements into avatar design, such as adding polarity indicators to the avatar's hands to illustrate the current direction in the circuit learning scenario. 
These observations indicate that while \textbf{our guidelines provide a solid foundation for designing intuitive avatars, additional scenario-specific guidance may be needed to fully support task-specific goals.} 
The guidelines could be broadened to include tailored design recommendations for different domains. 
In educational contexts, avatars need to include conceptual cues and instructional support. In contrast, avatars in gaming or social environments may focus more on narrative coherence, visual style, or identity expression.

\subsection{Supporting Multi-Capability Avatar Design and Adapting Existing Avatars}
Our guidelines are not limited to single-capability avatar design but also offer valuable support for multi-capability scenarios. 
First, designers can create multiple avatars in accordance with our guidelines, with each avatar representing a distinct augmented capability. 
These avatars can be integrated into a capability-switching system, enabling users to access different powers by switching between avatars. 

Second, \textbf{the guidelines also inform the design of multi-capability avatars by illustrating how different abilities can be represented through distinct visual elements or assigned to different body parts of the avatar. }
For example, we designed a flame pufferfish avatar that combines body size manipulation and fire manipulation (\autoref{Multicap}c). 
Following our design guidelines, the avatar adopts a pufferfish-like form to suggest its capacity to expand and contract (G1.1, G1.2, G3.1). 
Meanwhile, visual features such as red coloration (G1.6), molten rock textures (G1.4, G5.1), and black smoke emissions (G1.7) effectively communicate its association with fire-based powers.

Similarly, our guidelines can also be applied to modifying existing avatars. 
Designers can selectively apply the more detailed guidelines to modify the existing avatars, such as adding dynamic visual effects (G1.7) or attaching accessories (G1.8), to introduce new augmented capabilities. 
Guidelines related to inspiration or overall form may require more flexible interpretation. 
Designers can refer to these guidelines to modify specific body parts rather than changing the entire body shape. 
By adding distinct visual elements or modifying specific body parts of the avatar, designers can enhance existing avatars to represent new augmented actions while maintaining their original identity and visual consistency. 
Future work may further refine or adapt these guidelines to better support such modification workflows.

\subsection{Influence of Embodiment Perspective on Guideline Applicability}
While our guidelines are effective in helping create intuitive avatars that communicate augmented capabilities and corresponding interaction methods, their applicability can indeed be influenced by the user’s embodiment perspective. 
In fully immersive VR environments, users typically experience the world from a first-person view, which limits the visible range of their avatars to the hands, arms, or legs. 
However, virtual mirrors and avatar customization interfaces are common features in modern VR systems, allowing users to see their full-body avatars. 
For example, the Meta Quest platform provides a full-body avatar customization interface and a virtual mirror in its home environment. 
Therefore, \textbf{our guidelines can apply to most VR contexts where users can perceive their avatars beyond the first-person view. }
We recognize that in situations where only a first-person view is provided, additional design strategies may be necessary to ensure that augmented capabilities remain noticeable through the design of visible body parts. 
Conversely, in third-person VR experiences, our guidelines directly support avatar design, as augmented features and interaction cues are more fully observable.

\subsection{Practical Considerations for Adapting the Guidelines to VR System Design}
While our guidelines are effective in supporting the design of intuitive avatars, we acknowledge that applying them in real-world VR systems involves additional practical challenges. 
First, unintended gesture triggering is a common issue in immersive VR environments. Even with clear visual cues, users may accidentally activate functions due to natural body movements. 
Second, the effectiveness of avatar-based cues relies heavily on the precision of tracking and the responsiveness of the system. 
Inconsistent hand or body tracking or system latency can impact users' ability to perceive and perform the intended interactions correctly. Third, environmental and visual factors, including lighting, occlusion, and background complexity, can affect the visibility of avatar features. 
Specific visual indicators may become less noticeable in low-lighting conditions or when they blend with surrounding scene elements. 
These considerations suggest that \textbf{a complete VR implementation may require integrating additional interaction-level and system-level design strategies that account for technical and environmental constraints to guide intuitive avatar design in practice.}
Future work may explore extending our guidelines to explicitly address these practical challenges and provide structured design recommendations for robust deployment in real-world VR systems. 

\subsection{Broader Use of Avatar Design Guidelines}
Although our guidelines were developed in the context of VR avatars, their \textbf{applicability can extend beyond immersive environments, informing areas such as game character design and AI-assisted creative workflows. }
Many principles, such as visual metaphors, dynamic effects, and material cues to communicate capability and interaction, can inform character design in broader interactive media. 
For instance, in card-based or turn-based games with large character rosters, players often struggle to remember the abilities of each character. 
Applying intuitive visual design strategies from our guidelines could help convey key abilities at a glance, reducing cognitive load and improving usability. 
Similarly, in role-selection interfaces for competitive or narrative-driven games, visual expressiveness can support quicker recognition and enable more informed choices. 

Beyond assisting human designers, our guidelines may also serve as structured prompts for AI-based generative tools. 
With the rise of Large Multimodal Models (LMMs), designers can generate character concepts directly from text prompts. 
The design guidelines we proposed might steer generative processes toward more intentional and capability-focused avatar outcomes. 
This could accelerate the early design phase by producing more aligned starting points and could enable designers to iterate more effectively, refining, remixing, or adapting AI-generated results based on contextual needs.

\subsection{Limitations and Future Work}
While our guidelines supported the creation of intuitive avatars for augmented actions, we acknowledge that several limitations remain for future improvement. 
First, although our proposed classification covers a broad range of augmented actions, it may not represent the optimal or most comprehensive taxonomy. 
Some augmented actions may not neatly fit into our categories, but they could still benefit from the design strategies outlined in our guidelines. 
In future work, more augmented actions could be collected from a broader range of media and user-generated content to analyze, summarize, and expand the current classification. Additionally, parallel classification may be developed for other types of superpowers, such as passive attributes or sensory and cognitive enhancements. And the corresponding guidelines can also be extended, following the approach proposed in this work, to address additional types of augmented actions and other types of superpowers.  

Second, although we employed a structured process to develop the guidelines, the number of expert participants (N = 12) and avatar designs per action (2 to 3) were limited. 
This was primarily due to the challenge of recruiting qualified experts and the significant time investment required for each design, which took about two hours. 
These constraints may affect the generalizability of our findings. 
Additionally, cultural background and prior experiences may also influence how individuals perceive avatars and interpret their affordances. 
It may result in variations in how guidelines are applied or understood.  
However, the validation results from Study 2 indicate that the insights remain meaningful and representative. 
Future work could expand on this foundation by including a larger and more culturally diverse group of professional designers to further refine and validate the guidelines. 
Moreover, expanding the online survey to include a wider demographic range, such as children and older adults, would allow for a more effective capture of diverse perspectives and perceptual patterns. 

Third, participants in our study created static avatar sketches instead of fully modeled 3D avatars. 
While sketches are a standard early-stage format in avatar design, we acknowledge a gap between sketch-based design and full 3D implementation. 
To mitigate this limitation, we implemented four avatars in interactive VR applications for validation to address. 
Future work could explore using 3D avatar prototyping tools during the study to capture implementation constraints and assess design fidelity. 

Fourth, participants in the demonstration study expressed a clear preference for intuitive avatars over tutorials or textual instructions, because avatar-based cues allowed them to understand the capabilities and interaction methods more directly, reduced their learning effort, and made the functions easier to remember. 
For example, P7 noted, ``\textit{I can naturally associate the avatar with its capability and interaction. It lowers cognitive load.}'' 
However, our study primarily focused on how avatar design can intuitively convey augmented actions, rather than comparing it against other instructional approaches (e.g., text labels or tutorials) or exploring hybrid strategies.
Future research could compare different instructional modalities, such as avatars' visual cues and audio or textual instructions, to identify their strengths, weaknesses, and optimal combinations for effectively communicating augmented actions in VR. 

Fifth, the four demonstration scenarios were created to showcase the potential benefits of intuitive avatars in various VR contexts, rather than as finalized application designs. 
We recognize that these scenarios may not represent the most optimal use cases, and there may be more suitable applications. 
However, feedback from participants indicated that the intuitive avatars effectively enhanced their understanding and interaction within these scenarios. 
In real-world VR development, the process may be reversed.
Designers typically begin with a specific task or application context and then determine what augmented actions an avatar should support.  
Future research could explore more appropriate contexts and alternative application methods to better integrate intuitive avatars into diverse VR experiences. 

Finally, this study aimed to improve the intuitiveness of avatars, focusing on communicating augmented capabilities and interaction methods. 
As a result, we paid less attention to other aspects, such as visual appeal, aesthetic coherence, or emotional expressiveness. 
For example, P5 in the Demonstration Study described the lizard-inspired climbing avatar as ``\textit{slightly scary,}'' highlighting potential discomfort despite functional clarity. 
Additionally, our work focused primarily on visual design, without considering other modalities. 
Introducing additional modalities, such as haptic or auditory cues, could improve the intuitiveness and immersive quality of avatars. 
Future research could expand the guidelines using the elicitation approach proposed in this work to incorporate a broader range of design and multimodal considerations, enabling more holistic and context-sensitive avatar design across different domains.

%% file: sections/08_Conclusion.tex
\section{Conclusion}
In conclusion, we proposed a set of design guidelines for creating intuitive avatars that communicate augmented capabilities and corresponding interaction methods in VR. 
To ensure the generalizability of these guidelines, we first classified augmented actions into four categories and selected three representative actions from each, resulting in twelve in total. 
Twelve designers were invited to create avatars for these actions, focusing on conveying both the capability and corresponding interaction methods. 
Based on an analysis of their designs, we formulated sixteen guidelines, including eight general ones applicable across all actions and two specific to each category. 
To validate the effectiveness of the guidelines, we then conducted a controlled study in which two groups of eight participants designed avatars with and without the guidelines,  and twenty-four external judges evaluated the intuitiveness of the resulting avatars. 
The results showed that avatars created with our guidelines were significantly more effective in conveying the augmented capabilities and the associated interaction methods than those without.
Finally, we implemented four guideline-based avatars in VR applications across sports, navigation, and educational scenarios and invited twelve users to experience them. 
Their positive feedback indicated that the avatars enhanced immersion, engagement, and enjoyment across diverse VR contexts. 
We hope this work can inspire future research to further explore expressive and effective avatar design in virtual environments.